\documentclass[prd,preprint,showpacs,preprintnumbers,amsmath,amssymb,eqsecnum]{revtex4-1}

\usepackage{graphicx}
\usepackage{dcolumn}
\usepackage{bm}
\usepackage{subfigure}
\usepackage{color}

\begin{document}

\preprint{RUP-13-2}

\title{
Renormalization group approach to Einstein-Rosen waves
}
\author{$^{1}$Tomohiro Harada}%
 \email{harada@rikkyo.ac.jp}
\author{$^{2}$Sanjay Jhingan}
\email{sanjay.jhingan@gmail.com}
\affiliation{%
$^{1}$Department of Physics, Rikkyo University, Toshima, Tokyo 171-8501, Japan\\
$^{2}$
Centre for Theoretical Physics,
Jamia Millia Islamia, New Delhi 110025, India
}%
\date{\today}

\begin{abstract}
We present a renormalization group analysis to Einstein-Rosen waves or vacuum spacetimes with whole-cylinder symmetry.
It is found that self-similar solutions appear as fixed points in the renormalization group transformation. These
solutions correspond to the explosive gravitational waves and the
 collapsing gravitational waves at late times and
early times, respectively. Based on the linear perturbation analysis of the self-similar solutions, we conclude that
the self-similar evolution is stable as explosive gravitational waves under the condition of no incoming waves, while
it is weakly unstable as collapsing gravitational waves. The result implies that self-similar solutions can describe
the asymptotic behavior of more general solutions for exploding gravitational waves and thus extends the similarity
hypothesis in general relativity from spherical symmetry to cylindrical symmetry.
\end{abstract}


\maketitle



\section{Introduction}

Renormalization group analysis is a powerful tool for obtaining asymptotic behavior of solutions of partial
differential equations~\cite{Bricmont:1994}. Koike {\it et al.}~\cite{Koike:1995jm,Koike:1999eg} have applied this method to
a self-gravitating system in general relativity and successfully explained critical behavior together with
the critical exponent in gravitational collapse. Ib\'a\~nez and Jhingan~\cite{Ibanez:2004pt,Ibanez:2007yq}
have applied this method to inhomogeneous cosmology and analyzed the stability of scale-invariant asymptotic states
(see also Refs.~\cite{Chen:1994zza,Chen:1995ena,Iguchi:1997mx}).

Fixed points under renormalization group transformation are very important for the asymptotic analysis. The fixed-point solutions in general relativity generally correspond to self-similar solutions, which are defined as spacetimes that admit a homothetic Killing vector. Self-similar solutions arise from the scale-invariance of general relativity.
Carr~\cite{Carr:1993,Carr:2005uf} has conjectured that under certain physical circumstances, spherically symmetric
solutions will naturally evolve to a self-similar form from complicated
initial conditions, and this is termed as the similarity hypothesis. This
conjecture has been strongly supported in the gravitational collapse of a
perfect fluid by numerical simulation and linear stability analysis~\cite{Harada:2001nh}.

Although the similarity hypothesis has been proposed originally in
spherical symmetry, there is evidence that supports the validity of the
conjecture also in cylindrical symmetry. Nakao et
al.~\cite{Nakao:2009jp} have numerically simulated the collapse of the dust
cylinder and the subsequent emission of gravitational waves in
special cylindrical symmetry. 
Their numerical simulation suggests that gravitational waves gradually approach a self-similar form.

The vacuum Einstein equations in this symmetry can be analytically
integrated, and solutions are called the Einstein-Rosen
waves~\cite{Einstein:1937qu}. Since this system admits two commutative
spatial Killing vectors and still retains the dynamical degrees of freedom corresponding to gravitational waves, it provides us with a good toy model in which we can learn the physical nature of gravitational waves. Harada {\it et al.}~\cite{Harada:2008rx} have derived self-similar Einstein-Rosen waves and identified one of them with the attractor solution that Nakao {\it et al.}~\cite{Nakao:2009jp} reported.
In this paper, we recover the self-similar Einstein-Rosen waves
as a fixed point of the renormalization group analysis. Then, we perform the linearized perturbation analysis of these solutions (fixed
points)
and determine 
their  stability analytically using an eigenvalue analysis.

This paper is organized as follows. In Sec. II, we show the field
equations for Einstein-Rosen waves. In Sec. III, we introduce the
renormalization group analysis as a scaling transformation and
self-similar Einstein-Rosen waves as fixed points. In Sec. IV, we apply
linear perturbation theory to the self-similar solutions. In Sec. V, we
analyze the behavior of the perturbations at the boundaries. In Sec. VI, we show that the self-similar
Einstein-Rosen waves are stable and unstable at late times and at early times, respectively, under appropriate
boundary conditions. In Sec. VII, we conclude the paper. We use the units in which $c=1$.

\section{Einstein-Rosen waves}
We consider cylindrically symmetric spacetimes with the azimuthal Killing vector $\partial/\partial\phi$ and the translational Killing vector $\partial/\partial z$. We additionally assume that these two Killing vectors are
hypersurface orthogonal for {\it whole-cylinder symmetry}. Vacuum spacetimes with whole-cylinder symmetry
are called Einstein-Rosen waves~\cite{Einstein:1937qu}. The line element in Einstein-Rosen waves is given by
\begin{equation}
ds^{2}=e^{2(\gamma-\psi)}(-dt^{2}+dx^{2})+e^{-2\psi}x^{2}d\phi^{2}
+e^{2\psi}dz^{2}.
\end{equation}
The nontrivial Einstein equations take the following form:
\begin{eqnarray}
&& -\psi_{,tt}+\psi_{,xx}+\frac{1}{x}\psi_{,x}=0,
\label{eq:wave}\\
&& \gamma_{,x}=x(\psi_{,x}^{2}+\psi_{,t}^{2}),
\label{eq:gammax}\\
&& \gamma_{,t}=2x\psi_{,x}\psi_{,t}.
\label{eq:gammat}
\end{eqnarray}
We define new variables $h(x,t)$ and $f(x,t)$, which are more convenient to our analysis, by
\begin{equation}
f^{2}=\frac{h^{2}}{x^{2}}e^{2\gamma}\quad\mbox{and}\quad h^{2}=x e^{-2\psi},
\end{equation}
respectively.
Then, the line element is rewritten in the following form:
\begin{equation}
ds^{2}=f^{2}(-dt^{2}+dx^{2})+x(h^{2}d\phi^{2}+h^{-2}dz^{2}),
\end{equation}
where $f$ and $h$ satisfy
\begin{equation}
f_{,t}= \pm 2xf\frac{h_{,x}}{h}
 \left[{\frac{1}{4x^{2}}+\frac{1}{x}\frac{f_{,x}}{f}-\frac{h_{,x}^{2}}{h^{2}}}\right]^{1/2},
\label{eq:ft}
\end{equation}
and
\begin{eqnarray}
h_{,t}= \pm h
 \left[{\frac{1}{4x^{2}}+\frac{1}{x}\frac{f_{,x}}{f}-\frac{h_{,x}^{2}}{h^{2}}}\right]^{1/2},
\label{eq:ht}
\end{eqnarray}
respectively, where (and hereafter), in the case of double sign, we should
uniformly choose either an upper sign or lower sign.
The wave equation (\ref{eq:wave}) is identically satisfied from the
above two equations. Hence, we have the desired
evolution equations for the metric functions in a form suitable for
a renormalization group analysis.

\section{Renormalization group analysis}
Since we are interested in the solution in the asymptotic regime, we
shall explore now the scale-invariant properties of the system.
We consider the following scale transformation and define
the scaled functions, $h_{(L)}(x,t)$ and $f_{(L)}(x,t),$ as follows:
\begin{eqnarray}
x&\to& Lx, \\
t&\to & L^{\alpha}t ,\\
h&\to & L^{a}h(Lx,L^{\alpha}t)=h_{(L)}(x,t), \label{eq:scaling1}\\
f&\to & L^{b}f(Lx,L^{\alpha}t)=f_{(L)}(x,t). \label{eq:scaling2}
\end{eqnarray}

Here, $L$ is the scaling parameter, and scaled quantities $h_{(L)}$ and
$f_{(L)}$ satisfy the same evolution equations, i.e., Eqs.~(\ref{eq:ft})
and (\ref{eq:ht}).
From the structure of dynamical system, it is
easy to see that $t$ scales in the same way as $x$, fixing $\alpha = 1$.
There is no further constraint on scaling exponents $a$ and $b$.
Putting $t=1$ and redefining, in succession, $Lx$ by $x$ and $L$ by
$t$,  in Eqs.~(\ref{eq:scaling1}) and
(\ref{eq:scaling2}),
we find
\begin{eqnarray}
h(x,t)&=&t^{-a}h_{(L)}(x/t,1), \label{eq:recover_h}\\
f(x,t)&=&t^{-b}f_{(L)}(x/t,1)  \label{eq:recover_f}
\end{eqnarray}
in the scaling regime.
Note that the quantities $h_{(L)}(x/t,1)$ and $f_{(L)}(x/t,1)$ are
evaluated at some initial time ($t=1$ here) and are determined
as fixed points of the renormalization group equations as illustrated below.

\begin{center}

\begin{figure}
\includegraphics[width=0.3\textwidth]{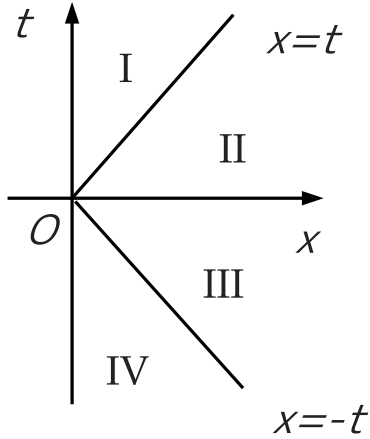}
\caption{\label{fg:domain} The light cone splits
the spacetime into four parts.
In the current paper, we focus on domains I and IV, i.e., the interior of the light cone.}
\end{figure}
\end{center}

In what follows, we adopt the mechanism developed by 
Bricmont {\it et al.}~\cite{Bricmont:1994}. Defining $L=\exp(\tau)$,
we can rewrite the scaling behavior of Eqs.~(\ref{eq:scaling1}) and
(\ref{eq:scaling2}) into a set of coupled ordinary differential equations:
\begin{eqnarray}
\frac{dh_{(L)}}{d\tau}&=&\left.ah_{(L)}+xh_{(L),x}+h_{(L),t}\right|_{t=1}, \\
\frac{df_{(L)}}{d\tau}&=&\left.bf_{(L)}+xf_{(L),x}+f_{(L),t}\right|_{t=1},
\end{eqnarray}
where the quantities on the right-hand side are evaluated at initial
time and are functions only of $x$.
The fixed-point solutions $h_{(L)}^{*}$ and $f_{(L)}^{*}$
are defined by
\begin{equation}
\frac{dh_{(L)}^{*}}{d\tau}=0, \quad \frac{df_{(L)}^{*}}{d\tau}=0.
\end{equation}
Therefore, we have fixed fixed points as solutions of a set of coupled nonlinear
ordinary differential equations:
\begin{eqnarray}
ah_{(L)}+xh_{(L),x}+
h_{(L)}\Psi=0,
\\
bf_{(L)}+xf_{(L),x}+ 2x f_{(L)}\frac{h_{(L),x}}{h_{(L)}}\Psi=0,
\end{eqnarray}
where $\Psi$ is defined as
\begin{equation}
 \Psi=\pm \left[\frac{1}{4x^{2}}+\frac{1}{x}\frac{f_{(L),x}}{f_{(L)}}-
 \left(\frac{h_{(L),x}}{h_{(L)}}\right)^{2}\right]^{1/2}.
\end{equation}
Here and henceforth, we have dropped superscript * from $h_{(L)}$ and $f_{(L)}$ for
brevity. These equations decouple on recombination, and we have the
following master equation for $h_{(L)}$:
\begin{equation}
\frac{h_{(L),x}}{h_{(L)}}=-\frac{a}{x}\pm
 \left[\frac{a^{2}+b-1/4}{x^{2}(1-x^{2})}\right]^{1/2}.
\label{eq:hLx}
\end{equation}
The other metric function $f_{(L)}$ can now be recovered as a solution
to an ordinary differential equation:
\begin{equation}
 \frac{f_{(L),x}}{f_{(L)}}=-\frac{b}{x}+\frac{2(a^{2}+b-1/4)}{x(1-x^{2})}
\mp \frac{2a\sqrt{a^{2}+b-1/4}}{x\sqrt{1-x^{2}}}.
\label{eq:fLx}
\end{equation}
We are interested in the domain $0<x<1$, which now corresponds to the
spacetime inside the light cone $0<x<t$. This restricts the parameter
space to $a^{2}+b-1/4>0$.
Equations~(\ref{eq:hLx}) and (\ref{eq:fLx}) can be easily integrated
to give the fixed points
\begin{eqnarray}
h_{(L)}&=&x^{-a\mp \Delta}(1-\sqrt{1-x^{2}})^{\pm\Delta},
\label{eq:hL}\\
f_{(L)}&=&c_{f}x^{-b+2\Delta^{2}\pm 2a\Delta}(1-x^{2})^{-\Delta^{2}}
(1-\sqrt{1-x^{2}})^{\mp 2a\Delta},
\label{eq:fL}
\end{eqnarray}
where $c_{f}$ is a constant of integration and $\Delta=a^{2}+b-1/4$.
Recovering time dependence through Eqs.~(\ref{eq:recover_h}) and
(\ref{eq:recover_f}),
we obtain
\begin{eqnarray}
h(x,t)&=&x^{-a\mp \Delta}(t-\sqrt{t^{2}-x^{2}})^{\pm\Delta}, \\
f(x,t)&=&c_{f}x^{-b+2\Delta^{2}\pm 2a\Delta}(t^{2}-x^{2})^{-\Delta^{2}}(t-\sqrt{t^{2}-x^{2}})^{\mp 2a\Delta}.
\end{eqnarray}
The light cone splits the spacetime domain $t>0$ 
into two parts, and we have another
solution corresponding to the region $x>t>0$ and $\Delta<0$. In fact,
Fig.~{\ref{fg:domain}} shows that the light cone splits the spacetime
into four parts if we consider both $t>0$ and $t<0$.
For the moment, we focus on domain I and later on IV, i.e., the
interior of the light cone. For domains II and III, i.e., the exterior
of the light cone, we have another solution corresponding to $x>|t|>0$
and $\Delta<0$, which is out of the scope of the present paper.

These solutions coincide with the self-similar or homothetic solutions
derived in Ref.~\cite{Harada:2008rx}, in which
the solutions are further restricted by imposing the solution
to be either with a regular axis  or a conically singular axis.
Then, the solutions are parametrized by two parameters $\kappa$ and
$\lambda$. The parameter $\kappa$ plays a more important role since it
determines the physical nature of the spacetime,
while $\lambda$ just controls the deficit angle of the axis.
The correspondence between $a$, $b$, and $c_{f}$ in the
above obtained expression and $\kappa$ and $\lambda$ in the expression
given in Ref.~\cite{Harada:2008rx}
is the following:
\begin{eqnarray}
\pm \Delta=\kappa, \quad a=\kappa-1/2, \quad b=\kappa, \quad c_{f}^{2}=2^{-4\kappa^{2}}e^{2\lambda},
\end{eqnarray}
where we should note that $\Delta\ge 0 $.
Note that we have started the analysis in the domain $0<x<t$.
Here, it should also be noted that both $\kappa=0$ and $1/2$
correspond to the flat spacetime, and $\kappa=-1/2$ corresponds to
the Kasner solution with exponents $(-1/2,2/3,2/3)$ or the 
locally rotationally symmetric Kasner solution.
The $x=t$ surface is a null singularity for $0<\kappa^{2}<1/4$ and
$1/4<\kappa^{2}<3/8$, a regular surface admitting
an extension beyond it for $3/8<\kappa^{2}<1/2$, 
and a null infinity for $\kappa^{2}\ge 1/2$.
We can further divide the case of $3/8<\kappa^2<1/2$ into the case
of $\kappa^{2}=1/2-1/(4n)$ ($n=2,3,\cdots$) and otherwise.
In the former, the $x=t$ surface admits an analytic extension
beyond it, while in the latter, this surface only admits a finitely
differentiable extension.
In general, these solutions describe
exploding gravitational waves. If we flip the sign of $t$,
these solutions describe the collapse of gravitational waves.
The fully detailed description of the physical nature and 
causal
structure of these solutions is given in Ref.~\cite{Harada:2008rx}.
The parameter value which fits the result of
the numerical simulation by Nakao {\it et al.}~\cite{Nakao:2009jp}
is $\kappa=-0.0206$,
for which the surface $x=t$ corresponds to a null singularity,
which is physically a shock of gravitational waves, and
the spacetime cannot be extended beyond it.

\section{Linear perturbation analysis}

We introduce the perturbation of the fixed point solution as follows:
\begin{eqnarray}
h_{(L)}&=&h_{(L)}^{*}(1+\delta h_{(L)}), \label{eq:deltah}\\
f_{(L)}&=&f_{(L)}^{*}(1+\delta f_{(L)}). \label{eq:deltaf}
\end{eqnarray}
Assuming $|\delta h_{(L)}|\ll 1$ and $|\delta f_{(L)}|\ll 1$,
we linearize the field equations for the perturbation.
The linearized perturbed equations take the form
\begin{eqnarray}
 \frac{d\delta h}{d\tau}&=&\left(x-\frac{1}{\Psi}\frac{h^{*}_{,x}}{h^{*}}\right)
\delta h_{,x}+\frac{1}{2x\Psi}\delta f_{,x},
\label{eq:ddeltahdtau}\\
 \frac{d\delta f}{d\tau}&=&
  2x\left[\Psi-\frac{1}{\Psi}\left(\frac{h^{*}_{,x}}{h^{*}}\right)^{2}\right]\delta
  h_{,x}+\left(x+\frac{1}{\Psi}\frac{h^{*}_{,x}}{h^{*}}\right)\delta f_{,x},
\label{eq:ddeltafdtau}
\end{eqnarray}
where we have dropped the subscript $(L)$ from the metric functions for
brevity.
Note that all the quantities are evaluated at the fixed points
(emphasized by an * ) and, hence, are
known functions of $x$ given by Eqs.~(\ref{eq:hL}) and (\ref{eq:fL}).

We now compute the normal modes with the dependence
$\delta h=e^{\omega\tau}\rho(x)$ and $\delta f=e^{\omega\tau
}\sigma(x)$, where $\omega$ is a constant. From Eqs.~(\ref{eq:ddeltahdtau})
and (\ref{eq:ddeltafdtau}),
we obtain
\begin{eqnarray}
 \omega\rho&=&\left(x+\frac{1}{x}+\frac{a}{x\Psi}\right)\rho_{,x}+
\frac{1}{2x\Psi}\sigma_{,x},
\label{eq:omegarho}\\
 \omega\sigma&=&\left[2\left(x\Psi-\frac{1}{x}\Psi-\frac{a^{2}}{x\Psi}\right)
-\frac{4a}{x}\right]\rho_{,x}+\left(x-\frac{1}{x}-\frac{a}{x\Psi}\right)
\sigma_{,x}.
\label{eq:omegasigma}
\end{eqnarray}
Combining the two equations, we can solve for $\sigma$ as
\begin{equation}
\omega\sigma=-2\omega(a\mp \Delta \sqrt{1-x^{2}})\rho\mp 2\Delta x\sqrt{1-x^{2}}\rho_{,x}. \label{eq:sigma}
\end{equation}
We can recover a master equation for $\rho(x)$ from
Eqs.~(\ref{eq:omegarho}), (\ref{eq:omegasigma}) and (\ref{eq:sigma}), i.e.,
\begin{equation}
\rho_{,xx}+\frac{2(\omega-1)x^{2}+1}{x(1-x^{2})}\rho_{,x}-\frac{\omega(\omega-1)}{1-x^{2}}\rho=0.
\label{eq:rho}
\end{equation}
Again, the above is estimated at $t=1$, and, hence,
$x$ should be regarded as $x/t$ in the notation of 
Harada {\it et al.}~\cite{Harada:2008rx}. Note that there is
no {\it a priori} reason to assume that $\omega$ is real.

We define $\zeta=1/\sqrt{1-x^{2}}$ and
$g=\zeta^{\omega}\rho$.
Then, Eq.~(\ref{eq:rho}) transforms to
\begin{equation}
\frac{d}{d\zeta}\left[(1-\zeta^{2})\frac{dg}{d\zeta}\right]+\omega(\omega+1)g=0,
\end{equation}
where $0<x<1$ corresponds to $\zeta>1$.
This is the Legendre differential equation. Note that this equation
has symmetry for the replacement of the parameter $\omega\leftrightarrow
-(\omega+1)$.
The Legendre functions of
the first and second kinds---$P_{\omega}(\zeta)$ and $Q_{\omega}(\zeta)$, respectively---are solutions of the Legendre
differential equation.
General solutions can be expressed in
terms of the Legendre functions simply as
\begin{equation}
\rho=(1-x^{2})^{\omega/2}\left[c_{1}P_{\omega}\left(\frac{1}{\sqrt{1-x^{2}}}\right)
+c_{2}Q_{\omega}\left(\frac{1}{\sqrt{1-x^{2}}}\right)\right].
\end{equation}
The above expression is convenient for ${\rm Re}(\omega)\ge -1/2$. For
${\rm Re}(\omega)<-1/2$, due to the symmetry of the Legendre differential
equation, we can express the solutions in terms of
the Legendre functions with indices $\nu=-(\omega+1)$ in the range ${\rm Re}(\nu)>-1/2$.
Then, we find the following expression that is more convenient:
\begin{equation}
\rho=(1-x^{2})^{\omega/2}\left[\bar{c}_{1}P_{-(\omega+1)}\left(\frac{1}{\sqrt{1-x^{2}}}\right)
+\bar{c}_{2}Q_{-(\omega+1)}\left(\frac{1}{\sqrt{1-x^{2}}}\right)\right].
\end{equation}
The solution for
another perturbation $\sigma$ of the metric functions can be
constructed by Eq.~(\ref{eq:sigma}).

\section{Behavior of the linear perturbations}
We impose the boundary conditions
from the physical requirements and study the allowed range of $\omega$.
To do this, we need the asymptotic behavior of $P_{\nu}(z)$ and
$Q_{\nu}(z),$ for $\mbox{Re}(\nu)\ge -1/2$ at $z=1$ and $z=\infty$, which
are given in Refs.~\cite{Erdelyi:1953,Virchenko:2001,Oldham:2008}.
The asymptotic behavior of $P_{\nu}(z)$ and $Q_{\nu}(z)$ at $z=1$
are given by
\begin{equation}
P_{\nu}(1)=1,
\end{equation}
and
\begin{equation}
Q_{\nu}(z)\simeq  -\frac{1}{2}\ln\frac{z-1}{2}-\gamma-\psi(\nu+1),
\quad (\nu\ne -1, -2, -3, \ldots,),
\end{equation}
 where
$\gamma$ is the Euler number and $\psi(z)=\displaystyle\frac{d}{dz}(\ln\Gamma(z))$ is the polygamma function.
The asymptotic behavior of $P_{\nu}(z)$ and $Q_{\nu}(z)$
for $z\to \infty$ is given by
\begin{eqnarray}
P_{\nu}(z)&\simeq& 2^{\nu}\pi^{-1/2}\Gamma(\nu+1/2)z^{\nu}/\Gamma(1+\nu),~ ({\rm Re}(\nu)>-1/2) \\
P_{-1/2}(z)&\simeq & \frac{\sqrt{2}}{\pi}\frac{\ln (8z)}{\sqrt{z}}, \\
P_{-1/2+ip}(z)&\simeq & \frac{i (2z)^{-1/2}}{\pi\tanh(\pi p)}
\left[\frac{\Gamma(1/2+i p)^{2}}{\Gamma(1+2i p)}(2z)^{-i p}-
\frac{\Gamma(1/2-i p)^{2}}{\Gamma(1-2ip)}(2z)^{i p}\right],
\end{eqnarray}
where $p\neq 0$, $p\in \mathbb{R}$ and 
\begin{equation}
Q_{\nu}(z)\simeq  2^{-\nu-1}\pi^{1/2}\Gamma(\nu+1)z^{-\nu-1}/\Gamma(\nu+3/2),
\end{equation}
respectively.

To make the discussion clear, we hereafter concentrate on the
stability of self-similar solutions with a regular or
only conically singular axis.
If we assume $\kappa=0$ for the parametrization of Harada
{\it et al.}~\cite{Harada:2008rx},
we find all the solutions of the family are flat.
Thus, we can assume $\kappa\ne 0$ or $\Delta\ne 0$
without the loss of generality.
The asymptotic behavior of the solutions at $x=0$ and $x=1$ are
given as follows.
\subsection{${\rm Re}(\omega)>-1/2$ and $\omega\ne 0$}
For ${\rm Re}(\omega)>-1/2$, we find
\begin{eqnarray}
\rho&\simeq& c_{1}+c_{2}[-\ln x+\ln 2 -\gamma-\psi(\omega+1)], \\
\sigma&\simeq& -2c_{1}(a\mp\Delta)+2c_{2}
[-(a\mp \Delta)[-\ln x+\ln 2 -\gamma-\psi(\omega+1)]\pm \Delta/\omega]
\end{eqnarray}
at $x=0$ and
\begin{eqnarray}
\rho&\simeq &
c_{1}2^{\omega}\pi^{-1/2}\frac{\Gamma(\omega+1/2)}{\Gamma(\omega+1)}+c_{2}2^{-\omega-1}\pi^{1/2}\frac{\Gamma(\omega+1)}{\Gamma(\omega+3/2)}(1-x^{2})^{\omega+1/2},
\\
\sigma&\simeq & -2c_{1}a 2^{\omega}\pi^{-1/2}\frac{\Gamma(\omega+1/2)}{\Gamma(\omega+1)}
\mp 2c_{2}\Delta 2^{-\omega-1}\pi^{1/2}\frac{(\omega+1/2)
\Gamma(\omega+1)}{\Gamma(\omega+3/2)}
(1-x^{2})^{\omega}
\end{eqnarray}
at $x=1$, where (and hereafter) only the relevant terms are shown.
Note that the gamma function has no zeroes.

\subsection{${\rm Re}(\omega)<-1/2$}
For ${\rm Re}(\omega)<-1/2$, we find
\begin{eqnarray}
\rho&\simeq& \bar{c}_{1}+\bar{c}_{2}[-\ln x+\ln 2 -\gamma-\psi(-\omega)], \\
\sigma&\simeq& -2\bar{c}_{1}(a\mp\Delta)+2\bar{c}_{2}
[-(a\mp \Delta)[-\ln x+\ln 2 -\gamma-\psi(-\omega)]\pm \Delta/\omega]
\end{eqnarray}
at $x=0$ and
\begin{eqnarray}
\rho&\simeq
&\bar{c}_{1}\frac{2^{-\omega-1}}{\pi^{1/2}}\frac{\Gamma(-\omega-1/2)}{\Gamma(-\omega)}(1-x^{2})^{\omega+1/2}+\bar{c}_{2}2^{\omega+1/2}\pi^{1/2}\frac{\Gamma(-\omega)}{\Gamma(-\omega+1/2)},
\\
\sigma&\simeq & -\bar{c}_{2}a 2^{\omega+3/2}\pi^{1/2}\frac{\Gamma(-\omega)}{\Gamma(-\omega+1/2)}
\mp \bar{c}_{1}\Delta
 \frac{2^{-\omega-1/2}}{\pi^{1/2}}\frac{(\omega+1/2)\Gamma(-\omega-1/2)}{\omega \Gamma(-\omega)} (1-x^{2})^{\omega}
\end{eqnarray}
at $x=1$.

\subsection{$\omega=0$}
For $\omega=0$, from Eq.~(\ref{eq:sigma}), we find $\rho=\mbox{const}$.
Substituting this into Eq.~(\ref{eq:omegasigma}), we find $\sigma=\mbox{const}$.
This corresponds to the rescaling of $t$ and $x$ by $At$ and $Ax$,
where $A$ is a positive constant. Hence, this is
not a physical mode but a gauge mode. In the following, we exclude this
zero mode from the analysis for this reason.

\subsection{$\omega=-1/2$}
For $\omega=-1/2$, we find
\begin{eqnarray}
\rho&\simeq& c_{1}+c_{2}[-\ln x+\ln 2 -\gamma-\psi(1/2)], \\
\sigma&\simeq& -2c_{1}(a\mp\Delta)+2c_{2}
[-(a\mp \Delta)[-\ln x+\ln 2 -\gamma-\psi(1/2)]\mp 2\Delta]
\end{eqnarray}
at $x=0$, and
\begin{eqnarray}
\rho&\simeq & \frac{\sqrt{2}}{\pi}c_{1}\ln \frac{8}{\sqrt{1-x^{2}}}
+\frac{\pi}{\sqrt{2}} c_{2}, \\
\sigma&\simeq & -\frac{2\sqrt{2}}{\pi}c_{1}\left(a\ln\frac{8}{\sqrt{1-x^{2}}}\mp
\frac{2\Delta}{\sqrt{1-x^{2}}}\right)-\sqrt{2}\pi c_{2}a
\end{eqnarray}
at $x=1$.
\subsection{$\omega=-1/2+ip$ ($p\ne 0$ and $p\in \mathbb{R}$)}
For $\omega=-1/2+ip$, we find
\begin{eqnarray}
\rho&\simeq& c_{1}+c_{2}[-\ln x+\ln 2 -\gamma-\psi(1/2+ip)], \\
\sigma&\simeq& -2c_{1}(a\mp\Delta)+2c_{2}
[-(a\mp \Delta)[-\ln x+\ln 2 -\gamma-\psi(1/2+ip)]\mp 2\Delta]
\end{eqnarray}
at $x=0$, and
\begin{eqnarray}
\rho&\simeq & c_{1}i\frac{2^{-1/2+ip}}{\pi \tanh(\pi p)}\left[
\frac{\Gamma(1/2+ip)^{2}}{\Gamma(1+2ip)} 2^{-2ip}(1-x^{2})^{ip}
-\frac{\Gamma(1/2-ip)^{2}}{\Gamma(1-2ip)}\right]
\nonumber \\
&& +c_{2}2^{-1/2-ip}\pi^{1/2}\frac{\Gamma(1/2+ip)}{\Gamma(1+ip)}(1-x^{2})^{ip}, \\
\sigma&\simeq & -\frac{2^{3/2-ip}}{-1/2+ip}\Delta \left[
\frac{c_{1}}{\pi \tanh(\pi p)}\frac{\Gamma(1/2+ip)^{2}}{\Gamma(1+2ip)}
+c_{2}\pi^{1/2} \frac{\Gamma(1/2+ip)}{\Gamma(1+ip)}\right](1-x^{2})^{-1/2+ip}
\end{eqnarray}
at $x=1$.

\section{Normal modes and stability analysis}

To see the stability in terms of normal modes, we impose boundary
conditions at $x=0$ and $x=1$ and see whether a growing
mode exits or not.
It is not trivial what boundary conditions we should impose on the solution. At least, the perturbation must be regular
at both points;
otherwise the linear perturbation scheme should break down.
Noting $a\mp \Delta=-1/2 $,
the regularity at $x=0$ requires $c_{2}=0$ for ${\rm Re}(\omega)\ge -1/2$
and $\bar{c}_{2}=0$ for ${\rm Re}(\omega) < -1/2$ because of the
logarithmic divergence of the Legendre function of the second kind at $x=0$.
We additionally impose regularity at $x=1$ and then find that
only ${\rm Re}(\omega) > -1/2$ is allowed. Conversely, for
${\rm Re}(\omega) > -1/2$ and $c_{2}=0$, the normal modes are
regular and analytic both at $x=0$ and $x=1$.
\subsection{Stability of the late-time solutions}
Our first motivation is to explain the numerical simulation
by Nakao {\it et al.}~\cite{Nakao:2009jp}. The parameter value
for the background is $\kappa=-0.0206$, for which the surface $x=t$ is
an outgoing null singularity, as is shown by Harada
{\it et al.}~\cite{Harada:2008rx}.
Since this null singularity is naked, we could inject
an incoming wave mode there, in principle.
However, the numerical simulation by Nakao {\it et al.}~\cite{Nakao:2009jp}
will correspond to no injection of such an incoming wave.
This corresponds to the condition that $\rho=\sigma=0$ at $x=1$. This, as well as the regularity at $x=0$, excludes all
normal mode perturbations.
This conclusion is valid also for all late-time ($t>0$) solutions
with any values of $\kappa$ under the condition of no incoming wave.
Note also that if we do not impose the no incoming wave condition but
only the regularity condition at $x=1$,
all normal mode perturbations with ${\rm Re}(\omega)>-1/2$ are 
allowed, and, hence, the self-similar evolution is strongly unstable.

\subsection{Stability of the early-time solutions}

So far, we have implicitly assumed late-time solutions,
i.e., $t>0$. However, it is also interesting to study the stability of
early-time ($t<0$) solutions
in the context of gravitational collapse.
The early-time solutions can be obtained by just flipping
the sign of $t$. Since in this case $x=-t$ is an
ingoing null singularity or regular surface,
we can impose the regularity at $x=0$ and $x=1$.
We have seen that under the regularity at both $x=0$ and $x=1$, 
only ${\rm Re}(\omega) > -1/2$ is allowed.
Since the physical time evolution from $t=-\infty$ to $t=0$
corresponds to the decrease of $\tau$ from $\tau=\infty$ to $\tau=-\infty$,
we can conclude that the self-similar early-time solutions
are weakly unstable inside the null surface $x=-t$
against the normal modes with eigenvalue $\omega$ satisfying
$-1/2<{\rm Re}(\omega) < 0 $.
If ${\rm Re}(\omega) > -1/2$, all the values for $\omega$ are allowed.
However, during the time evolution, the most rapidly growing mode
will dominate the perturbation. In this sense, growing modes
with the growth rate ${\rm Re}(\omega) = -1/2+0$ will dominate the
perturbation, and this suggests that the critical exponent is
2 if there appears a scaling law for the quantity
of the mass dimension. However,
there exist a countably infinite number of unstable modes;
critical behavior in this system would be very different
from those which have been studied so far.
In the present context, we do not need to refer to the behavior
of the perturbation outside the surface $x=-t$.
This is consistent with the
renormalization group analysis for the spherically
symmetric gravitational collapse, initiated by
Koike {\it et al.}~\cite{Koike:1995jm,Koike:1999eg}.
If we additionally impose the condition $\rho=\sigma=0$ at $x=1$ as in
the late-time case, all normal mode perturbations are excluded, 
and, hence, the self-similar evolution becomes stable.

\section{Conclusion}

The numerical simulation by Nakao {\it et al.}~\cite{Nakao:2009jp}
strongly suggests
that a self-similar gravitational wave acts as an attractor
in the vacuum region at late times
after the explosive burst of cylindrically symmetric
gravitational radiation in whole-cylinder symmetry.
Motivated with this numerical result, we study the stability of
self-similar Einstein-Rosen waves.
There are exact solutions that describe the self-similar
Einstein-Rosen waves. The parameter values are identified for
the numerical solution found in Ref.~\cite{Nakao:2009jp}, in which
the null surface $x=t$ corresponds to a null singularity
or a shock gravitational wave.

We have analyzed the behavior of cylindrically
symmetric linear perturbation
around the self-similar solutions in terms of the normal mode
analysis. We have found that the late-time self-similar solutions are
stable inside the surface $x=t$ of gravitational waves
at late times under the no incoming wave condition.
This is the case for all the self-similar Einstein-Rosen waves
at late times.

We have also investigated the stability of self-similar
Einstein-Rosen waves at early times, which describe
the collapse of gravitational waves. This is important in the
context of gravitational collapse.
We find that self-similar Einstein-Rosen waves are weakly
unstable inside the surface $x=-t$ against regular
cylindrically symmetric perturbations.

The former case provides us with the demonstration that
self-similar solutions that arise from
the scale invariance of general relativity play
an important role in the dynamics of gravitational waves
and, hence, extends the applicability
of the similarity hypothesis~\cite{Carr:2005uf},
which was originally proposed for spherically
symmetric spacetimes.
We should note that the locally rotationally symmetric Kasner solution 
is known to be unstable against some sorts of homegenous but anisotropic 
perturbations~\cite{Wainwright_Ellis:1997}. This suggests that 
the self-similarity hypothesis is a consequence of the restriction to 
the models with few enough degrees of freedom, which, hence, stabilize 
the self-similar solutions that are not stable in general.
The renormalization group analysis together with linear perturbation
scheme is quite useful to understand the asymptotic behavior of the
somewhat general solutions of partial differential equations.

\begin{acknowledgements}
T.H. thanks T. Koike, K. Nakamura and K.~I. Nakao
for fruitful discussion. T.H. was supported by
Grant-in-Aid for Scientific Research from
the Ministry of Education, Culture, Sports, Science and Technology of
Japan [Young Scientists (B) Grant No. 21740190].
T.H. thanks Centre for Theoretical Physics, Jamia Millia Islamia, for
its hospitality. S.J. thanks J. Ib{\'a}{\~n}ez for discussions and
 gratefully acknowledges support from the Rikkyo University Research
 Associate Program. S.J. acknowledges support under a UGC 
 minor research project on black holes and visible singularities. 
 T.H. and S.J. would like to thank the anonymous referee for helpful comments.
\end{acknowledgements}


\end{document}